\documentclass[aps,preprint]{revtex4}%
\usepackage{amsfonts}
\usepackage{amsmath}
\usepackage{amssymb}
\usepackage{graphicx}%
\providecommand{\U}[1]{\protect\rule{.1in}{.1in}}

\ifx\pdfoutput\relax\let\pdfoutput=\undefined\fi
\newcount\msipdfoutput
\ifx\pdfoutput\undefined\else
\ifcase\pdfoutput\else
\ifx\paperwidth\undefined\else
\ifdim\paperheight=0pt\relax\else\pdfpageheight\paperheight\fi
\ifdim\paperwidth=0pt\relax\else\pdfpagewidth\paperwidth\fi
\fi\fi\fi
\begin{document}
\title{Conicoid Mirrors}
\author{Diego Casta\~{n}o}
\affiliation{Nova Southeastern University}
\author{Lawrence Hawkins}
\affiliation{Nova Southeastern University}
\keywords{optics, mirrors}
\begin{abstract}
The first order equation relating object and image location for a mirror of
arbitrary conic-sectional shape is derived. It is also shown that the
parabolic reflecting surface is the only one free of aberration and only in
the limiting case of distant sources.

\end{abstract}
\date{September 2010}
\startpage{1}
\endpage{ }
\maketitle

\section{\bigskip Introduction}

Most elementary treatments of reflecting surfaces restrict their attention to
the spherical case. In this standard case, and assuming the paraxial
approximation (all angles are small and all rays are close to the optical
axis), the resulting equation relating the \emph{axial} object and image
positions and the radius of curvature of the reflecting spherical surface is
\begin{equation}
\frac{1}{v}+\frac{1}{u}=\frac{2}{r}, \label{gauss}%
\end{equation}
where all parameters are one dimensional coordinates which locate the image
($v$), object ($u$), and center of curvature ($r$) with respect to the vertex
(the intersection of the surface with the optical axis) \cite{pedrotti}. A
convention is typically assumed in which light rays travel from left to right
in all figures. The origin of the one dimensional coordinate system employed
coincides with the vertex, and locations to the right (left) of the vertex are
positive (negative).
\begin{figure}
[ptbh]
\begin{center}
\includegraphics[scale=0.75]{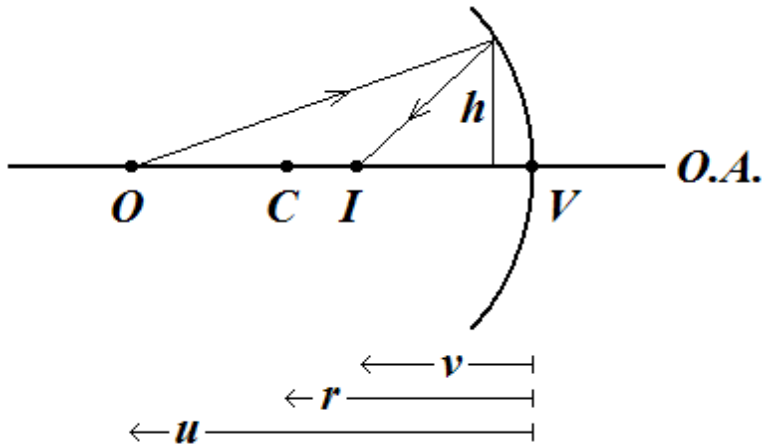}%
\caption{The spherical reflecting surface. }%
\label{fig01}%
\end{center}
\end{figure}
The paraxial approximation is equivalent to a first order approximation in the
height ($h$) of the incidence point (on the surface) of a reflecting ray. To
higher order, it is found that
\begin{equation}
v=v(u,r,h). \label{v(h)}%
\end{equation}
Consequently, spherical mirrors are aberrant at higher order since the image
location is not independent of the height, $h$.

This paper represents a more general treatment of a mirror than is typically
found in the literature. The reflecting surface is assumed to be a conicoid,
the surface of revolution generated by a conic. Equation (\ref{gauss}) is then
derived as the special case of a spherical surface and to first order in $h$.
Special cases are analyzed as a function of asphericity, or departure from the
spherical, of the reflecting surface. The parabolic surface is shown to be
uniquely special in that
\begin{equation}
\frac{dv}{dh}=0 \label{dvdh}%
\end{equation}
to all orders for objects at infinity ($u\rightarrow-\infty$).

\section{Analysis: Conicoid Case}%

\begin{figure}
[ptbh]
\begin{center}
\includegraphics[scale=0.75]{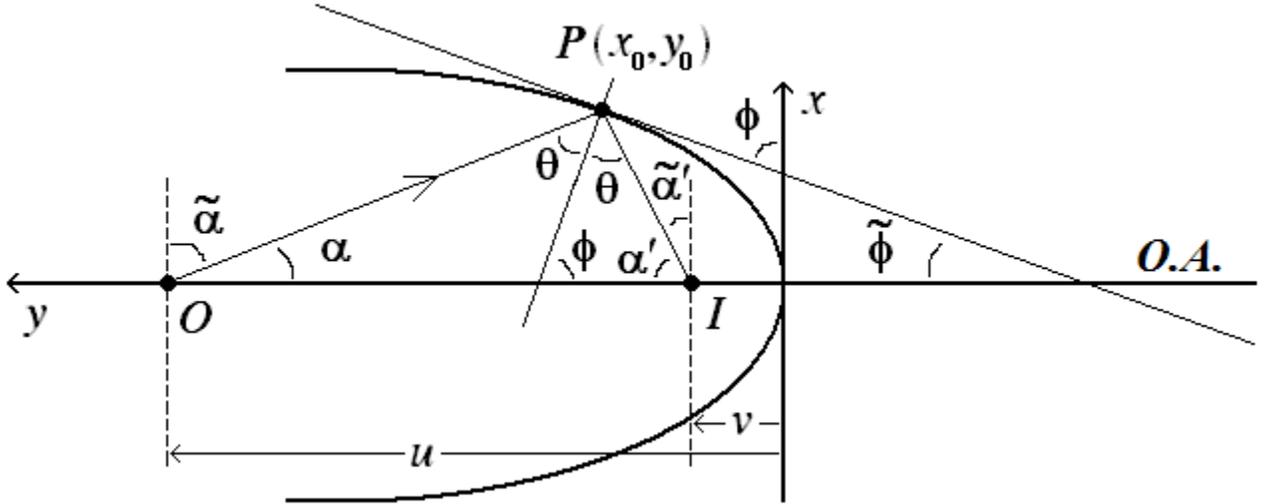}
\caption{Conicoid mirror.}%
\label{fig02}%
\end{center}
\end{figure}

In Fig. \ref{fig02}, a conicoid reflecting surface is depicted with equation
\begin{equation}
x^{2}=\frac{1}{a}y-\sigma y^{2},\;a=-\frac{1}{2r}>0, \label{eq01}%
\end{equation}
where $r$ is the radius of curvature of the surface at the vertex, and
$\sigma=1-e^{2}$ is the shape factor and is related to the standard
eccentricity (see Appendix I or, for example, \cite{conicoid}). For a sphere,
$\sigma=1$, whereas for a paraboloid $\sigma=0$. Note that the $xy$ coordinate
system is set on its side so that $+\hat{y}$ coincides with the negative
direction on the optical axis (O.A.) as defined in Fig. \ref{fig01}\ of the
Introduction. Consequently, the radius of curvature, $r$, at the origin for
any concave conicoid (\textit{i.e.}, opening to the left) will be considered
negative. In Fig. \ref{fig02}, a representative case is depicted with $u<0$,
the location of the object, and $v<0$, the location of the image. The figure
displays an incident ray, $\overline{OP}$, emanating from the object at $O$
and a reflected ray, $\overline{IP}$, passing through the image at $I$. From
the figure, the line $\overline{OP}$ has equation in the $xy$-plane
\begin{equation}
y-(-u)=x\tan\left(  -\tilde{\alpha}\right)  =-\frac{x}{\tan\alpha}.
\label{eq02}%
\end{equation}
Similarly, the line $\overline{IP}$ has equation
\begin{equation}
y-(-v)=x\tan\tilde{\alpha}^{\prime}=\frac{x}{\tan\alpha^{\prime}}.
\label{eq03}%
\end{equation}
Consequently,
\begin{align}
\tan\alpha &  =-\frac{x_{0}}{y_{0}+u},\label{eq04a}\\
\tan\alpha^{\prime}  &  =\frac{x_{0}}{y_{0}+v}, \label{eq04b}%
\end{align}
where $(x_{0},y_{0})$ is the point of reflection, $P$, on the surface. From
the figure, it follows that
\begin{equation}
\phi=\alpha+\theta, \label{eq05}%
\end{equation}
where
\begin{equation}
\tan\phi=\left.  \frac{dy}{dx}\right\vert _{P}=\frac{2ax_{0}}{1-2\sigma
ay_{0}}, \label{eq06}%
\end{equation}
and
\begin{equation}
\alpha^{\prime}+\theta+\phi=\pi. \label{eq07}%
\end{equation}
Therefore
\begin{align}
\tan\left(  \alpha-\alpha^{\prime}\right)   &  =\tan\left(  2\phi-\pi\right)
\label{eq08a}\\
\frac{\tan\alpha-\tan\alpha^{\prime}}{1+\tan\alpha\tan\alpha^{\prime}}  &
=\frac{2\tan\phi}{1-\tan^{2}\phi}. \label{eq08b}%
\end{align}
Substituting for the tangents from above yields
\begin{equation}
\left(  \frac{1}{v}+\frac{1}{u}\right)  \left[  1-\frac{4\sigma a^{2}y_{0}%
^{2}}{\left(  1-2\sigma ay_{0}\right)  ^{2}}\right]  -\left(  \frac{1}%
{uv}\right)  \frac{2y_{0}\left[  1+2\left(  1-\sigma\right)  ay_{0}\right]
}{\left(  1-2\sigma ay_{0}\right)  ^{2}}=-\frac{4a}{1-2\sigma ay_{0}}
\label{eq09a}%
\end{equation}%
\begin{align}
\left(  \frac{1}{v}+\frac{1}{u}\right)  \left[  1-4\sigma ay_{0}%
-4\sigma\left(  1-\sigma\right)  a^{2}y_{0}^{2}\right]  -\left(  \frac{1}%
{uv}\right)   &  2y_{0}\left[  1+2\left(  1-\sigma\right)  ay_{0}\right]
\hspace{0.45in}\nonumber\\
&  \hspace{0.6in}=-4a\left(  1-2\sigma ay_{0}\right)  . \label{eq09b}%
\end{align}

Now let $h\equiv x_{0}$ be the height of the incidence point $P$ for a
particular ray from the source object at $O$, then in the paraxial
approximation ($h\ll1$),
\begin{equation}
y_{0}=ah^{2}+\sigma a^{3}h^{4}+\mathcal{O}(h^{6}). \label{eq10}%
\end{equation}
Equation (\ref{eq09b}) can then be rewritten to fourth order as
\begin{align}
\left(  \frac{1}{v}+\frac{1}{u}-\frac{2}{r}\right)   &  =\left[  2a\left(
\frac{1}{uv}\right)  -8\sigma a^{3}\right]  h^{2}\nonumber\\
&  \hspace{0.5in}+\left[  4\sigma a^{4}\left(  \frac{1}{v}+\frac{1}{u}\right)
+2\left(  3\sigma+2\right)  a^{3}\left(  \frac{1}{uv}\right)  -24\sigma
^{2}a^{5}\right]  h^{4}. \label{eq11}%
\end{align}
Note that there is aberration in imaging a finite axial point since there is
no confluence in the rays from $O$. Also note that there is no fixed shape
factor $\sigma$ that eliminates aberration to second order and higher. To
first order, all conicoids obey the same relation
\begin{equation}
\frac{1}{v}+\frac{1}{u}=\frac{1}{f}+\mathcal{O}(h^{2}), \label{eq12}%
\end{equation}
which coincides, of course, with the Gaussian (first order approximation)
equation for a spherical mirror with focal length $f=r/2$.

From Eq. (\ref{eq09b}) it follows that for objects at infinity ($u\rightarrow
-\infty$) and a parabolic shape ($\sigma=0$), the image forms at $v=f$
regardless of the height of the incidence ray, therefore, there is no
aberration for such imaging.

\section{Analysis: General Case}

It is desirable to know to what extend the results of the previous section are
pathological to conicoids. With this in mind consider the most general
axi-symmetric surface of revolution (about the y-axis) as a reflector%

\begin{equation}
y=\sum_{n=1}^{\infty}c_{2n}x^{2n}. \label{eq13}%
\end{equation}
Equation (\ref{eq08b}) is easily generalized to
\begin{equation}
\frac{\tan\alpha-\tan\alpha^{\prime}}{1+\tan\alpha\tan\alpha^{\prime}}%
=\frac{2y^{\prime}}{1-y^{\prime2}}, \label{eq14}%
\end{equation}
where $y^{\prime}=\left.  \frac{dy}{dx}\right\vert _{P}$. In general, for a
given axial object location, the image location (or intersection point of the
reflected ray with the optical axis) is a function of the object location and
the reflection point
\begin{equation}
v=v(u,x_{0})\text{ with }y_{0}=y(x_{0}). \label{eq15}%
\end{equation}
A reflecting surface is free of aberration if
\begin{equation}
v^{\prime}\equiv\frac{dv}{dx_{0}}=0\;\forall u. \label{eq16}%
\end{equation}
Equation (\ref{eq14}) can be implicitly differentiated to yield
\begin{align}
\frac{v^{\prime}}{v}\left\{  2y^{\prime}+\frac{x-xy^{\prime2}+2yy^{\prime}}%
{u}\right\}   &  =\left(  \frac{1}{v}+\frac{1}{u}\right)  \left\{
\frac{\left(  1+y^{\prime2}\right)  \left(  xy^{\prime\prime}-y^{\prime
}\right)  }{y^{\prime}}\right\}  _{1}\nonumber\\
&  \;\;\;\;+2\left(  \frac{1}{uv}\right)  \left\{  \frac{\left(  1+y^{\prime
2}\right)  \left[  \left(  xy^{\prime\prime}-y^{\prime}\right)  y+xy^{\prime
2}\right]  }{y^{\prime}}\right\}  _{2}. \label{eq17}%
\end{align}
The aberration-free surface must satisfy $\left\{  \cdots\right\}
_{1}=\left\{  \cdots\right\}  _{2}=0$. However, it is evident from Eq.
(\ref{eq17}) that this cannot be obtained trivially. For the special case in
which the object is at infinity though, the aberration-free surface must only
satisfy $\left\{  \cdots\right\}  _{1}=0$, and this leads to a defining
equation for the surface
\begin{equation}
xy^{\prime\prime}-y^{\prime}=0. \label{eq18}%
\end{equation}
This is a linear differential equation whose general solution can most easily
be found by the reduction in order method to give the general solution
$y=Ax^{2}+B$. This further reduces to the particular solution of Eq.
(\ref{eq32}), found by another method, after the two needed boundary
conditions are invoked.

\section{Conclusion}

Most elementary treatments of mirrors lack a discussion of the first order
equation relating object and image locations in the case of arbitrary mirror
shape. The default reflecting surface is always the spherical one. In fact, a
simple analysis yields that all axi-symmetric, conic, reflecting surfaces of
revolution (conicoids) in the first order, paraxial approximation satisfy the
same (Gaussian) equation
\begin{equation}
\frac{1}{v}+\frac{1}{u}=\frac{2}{r}, \label{eq28}%
\end{equation}
where $r$ is the radius of curvature of the surface at its vertex.

Aberrations enter at second order and cannot be eliminated for finite object
locations by any fixed shape. However, for objects at infinity, or
specifically, for incoming light parallel to the optical axis, there is a
unique reflecting shape that is free of aberration -- the parabolic one.

\appendix

\section{Derivation of the Conic Section Equation}

Starting with the general form of a conic section in Cartesian coordinates,
\begin{equation}
Ax^{2}+Bxy+Cy^{2}+Dx+Ey+F=0, \label{eq19}%
\end{equation}
assume $x$-reflection symmetry, so that the equation reduces to
\begin{equation}
x^{2}+C^{\prime}y^{2}+E^{\prime}y+F^{\prime}=0. \label{eq20}%
\end{equation}
Next the curve is shifted so the vertex coincides with the origin,
$y\rightarrow y+k,$ with $2C^{\prime}k+E^{\prime}=\pm\sqrt{E^{\prime
2}-4C^{\prime}F^{\prime}}$. If the form is further constrained so that the
curve lies in $y\geq0$ half-plane, then the positive root is required, and
this yields
\begin{equation}
x^{2}=\sqrt{E^{\prime2}-4C^{\prime}F^{\prime}}y-C^{\prime}y^{2}, \label{eq21}%
\end{equation}
or in terms of new parameters
\begin{equation}
x^{2}=\frac{1}{a}y-\sigma y^{2}, \label{eq22}%
\end{equation}
where $a>0$. The signed curvature of this curve at the origin is
\begin{equation}
k\equiv\left.  \frac{y^{\prime\prime}}{\left(  1+y^{\prime2}\right)  ^{3/2}%
}\right\vert _{(0,0)}=2a>0. \label{eq23}%
\end{equation}
Given the optics conventions adopted here as described in the Introduction and
depicted in Figures 1 and 2, the radius, $r$, of the osculating circle at the
origin for a concave conicoid is considered negative. The radius of curvature
is therefore related to the parameter $a$
\begin{equation}
k=2a=-\frac{1}{r}>0, \label{eq24}%
\end{equation}
and
\begin{equation}
\frac{1}{a}=-2r. \label{eq25}%
\end{equation}

From Eq. (\ref{eq22}) it follows that $\sigma=0$ corresponds to a parabola. By
putting Eq. (\ref{eq22}) into canonical form
\begin{equation}
\frac{x^{2}}{\left(  \frac{1}{4\sigma a^{2}}\right)  }+\frac{\left(
y-\frac{1}{2\sigma a}\right)  ^{2}}{\left(  \frac{1}{4a^{2}}\right)  }=1,
\label{eq26}%
\end{equation}
it becomes clear that $\sigma=1$ corresponds to a circle with radius
$\left\vert r\right\vert =\frac{1}{2a}$. The equation describes a hyperbola
when $\sigma<0$. For $0<\sigma<1$, the equation describes an oblate ellipse
(with respect to the y-axis), and it describes a prolate ellipse for
$\sigma>1$. In fact, from Eq. (\ref{eq26}), the shape factor, $\sigma$, can be
related to the standard eccentricity
\begin{equation}
\sigma=1-e^{2}. \label{eq27}%
\end{equation}

\section{Alternate Derivation of the Paraboloid in the Limiting Object
Distance Case}

\begin{figure}
[ptbh]
\begin{center}
\includegraphics[scale=0.75]{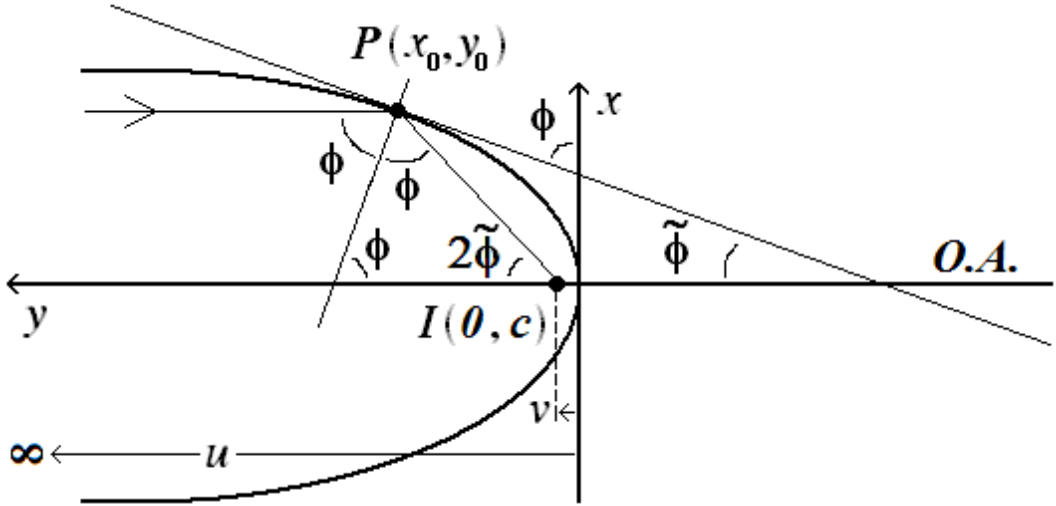}%
\caption{Paraxial incident ray on unknown conicoidal surface. }%
\label{fig03}%
\end{center}
\end{figure}

An alternate solution (to that of Section III) is presented for the exact
conicoid shape in the limit that the object distance approaches infinity
($u\rightarrow\infty$). Applying the law of reflection (based on Fermat's
principle of stationary optical path) to a parallel (to the optical axis) ray
(from a distant object) incident on an unknown conicoid surface, results in
the optical path displayed in Fig. \ref{fig03}.
Applying Eq. (\ref{eq04b}) to the present special case, it follows that
\begin{equation}
\tan2\tilde{\phi}=\frac{2\tan\tilde{\phi}}{1-\tan^{2}\phi}=\frac{x_{0}}%
{y_{0}-c}. \label{eq29}%
\end{equation}
If the notation is changed and the $y$ variable is shifted for convenience,
$(x_{0},y_{0})\rightarrow(x,y+c),\;\tan\tilde{\phi}=\frac{dx}{dy}$, then Eq.
(\ref{eq29}) can be reduced to either a homogeneous nonlinear ordinary
differential equation (ODE) of the form
\begin{equation}
\frac{dx}{dy}=\frac{-y+k\sqrt{x^{2}+y^{2}}}{x},\;k=1,\;x(-c)=0 \label{eq30}%
\end{equation}
or to a nonlinear Clairaut ODE \cite{clairaut} of the form
\begin{equation}
w=y\frac{dw}{dy}+\frac{1}{4}\left(  \frac{dw}{dy}\right)  ^{2},\;w=x^{2}.
\label{eq31}%
\end{equation}
Recall that Clairaut solutions are of the form $w(y)=my+f(m)$ and have
envelopes that are also exact singularity solutions. Solving Eq. (\ref{eq30})
or (\ref{eq31}) yields the final form for the unknown conicoid (and shifting
back $y\rightarrow y-c$)
\begin{equation}
y=\frac{1}{4c}x^{2}, \label{eq32}%
\end{equation}
which is the equation for the (meridional) cross section of a paraboloid with
focus at $(0,c)$. It is also of note that Eq. (\ref{eq30}) with $k>0$ can be
used to model various and sundry airplane, ship, and predator/prey pursuit
problems \cite{pursuit}.


\begin{thebibliography}{9}                                                                                                %


\bibitem {pedrotti}L. Pedrotti and F. Pedrotti, \textit{Optics and Vision} (Prentice-Hall,1998).

\bibitem {conicoid}J.W. Foreman, "The Conic Sections Revisited," Am. J. Phys.
\textbf{59}, 1002-1005 (1991); D.M. Watson, \textit{Astronomy 203/403:
Astronomical Instruments and Techniques On-Line Lecture,} University of
Rochester (1999), \textbf{http://www.pas.rochester.edu/\symbol{126}%
dmw/ast203/Lectures.htm.}

\bibitem {clairaut}R. K. Nagle, E. B. Saff, and A. D. Snider,
\textit{Fundamentals of Differential Equations}, 7th Edition (Pearson /
Addison Wesley, 2008), pp. 88-89 \qquad\qquad\qquad\qquad\qquad\qquad
\qquad\qquad\qquad

\bibitem {pursuit}Dennis Zill, \textit{Differential Equations with modeling
applications} (Brooks/Cole, 2005), p. 123-125; G. F. Simmons and S. G. Krantz,
\textit{Differential Equations} (McGraw-Hill 2007), pp.42-45.
\end{thebibliography}
\end{document}